\documentclass[journal,comsoc]{IEEEtran}

\IEEEoverridecommandlockouts
\usepackage{cite}
\usepackage{graphicx}
\usepackage{mdwmath}
\usepackage[font={small}]{caption}

\begin{document}

\author{Soroush~Khaleghi,~\IEEEmembership{Member,~IEEE,}
      Wenjing~Rao,~\IEEEmembership{Member,~IEEE,}}

\title{Repairability Enhancement of Scalable Systems with Locally Shared Spares
}
\maketitle

\begin{abstract}
Future systems based on nano-scale devices will provide great potentials for scaling up in system complexity, yet they will be highly susceptible to operational faults. While spare units can be generally used to enhance reliability, they must be shared in a limited way among functional units to ensure low-cost overheads when systems scale up. 
Furthermore, the efficiency of achieving reliability using spare units heavily depends on the replacement mechanisms of such spares. While global and chained replacement approaches can take advantage of the entire replacement capabilities in the network, they usually impose some sort of disturbance to all the functional units in the system during the repair process, thus are dreadfully expensive in terms of performance overhead for systems with high fault rates. In this paper, we focus on a low-cost, fast, ``immediate" replacement mechanism that can be implemented locally with minimum disturbance to the system.  
The proposed schemes aim for maintaining the system with high fault rates in such a low-cost, fast repairable status for many faults before invoking the more expensive, yet optimal, approaches.
First, we propose an online repair algorithm: as faults occur during the run-time of the system, the proposed algorithm makes a choice of a spare unit (among several candidates), such that the overall impact on system repairability in the future is minimized. Second, we propose a network enhancement approach, which identifies and connects the vulnerable units to the exploitable spares, thus strengthening the entire system at a low cost.

\end{abstract}

\section{Introduction}
\label{intro}
As feature sizes continue to shrink with the advances in nanotechnologies, future systems are expected to scale up in complexity and computational capabilities. However, the minute scale of nano devices makes them highly susceptible to operational faults \cite{road} \cite{Beckett} \cite{Nanofault}. Therefore, one of the critical limitations facing all future systems (based on the emerging nanoelectronic devices or the progressively scale-down CMOS ones) is unreliability. 

However, high reliability is an absolute necessity in many applications, such as robotic systems, medical devices, and transportation systems \cite{du1997degradable} \cite{parastegari2016fail} \cite{christensen1985adherence}. Therefore, Hardware redundancy techniques such as \textit{fault masking} and \textit{repair-based} schemes are typically employed to boost reliability. 

In the fault masking approaches, such as N-Modular Redundancy \cite{NMR1}, multiple copies of a component perform the same task to produce a single output through a majority vote. These schemes are deemed very expensive as fault rates grow, as is the case of nano systems \cite{NMR3}. Repair-based approaches, on the other hand, rely on fault detection and standby spare units. After detecting the presence of a fault, a subsequent repair process replaces the faulty unit with a standby spare unit \cite{spare1} \cite{spare2} \cite{spare3} \cite{spare4} \cite{banerjee2015conditions} \cite{banerjee2017local} \cite{DFTpaper} \cite{khaleghi2013constructing}. Since each spare unit can be used to replace several functional units, the repair-based approaches usually require smaller hardware overheads. Furthermore, as the energy efficiency of devices does not scale along with the integration capacity, the area of the chip that cannot be powered has been increasing. This trend, known as \emph{Dark Silicon}  \cite{dark}, makes it more plausible to employ repair-based schemes with standby spares in future systems.

Repair-based approaches become most efficient when spare units can be shared among several functional units, i.e., each spare unit can be used to replace any of multiple functional units \cite{NMR3} \cite{repair2}. 
Such a ``spare sharing" approach requires the functionality of a spare unit to be compatible with those of the functional units to be replaced. 
%For instance, a spare Ripple Carry Adder (RCA) might be used to replace another RCA or even a Carry Look-Ahead Adder (at the cost of lower speed), yet it cannot replace a faulty multiplier. 
Such a compatibility is either ensured by employing redundant units, or via reconfigurability, for example in the case of Field Programmable Gate Arrays (FPGAs) and Multi-Processor System-on-Chips (MPSoCs) \cite{banerjee2015conditions} \cite{banerjee2017local} \cite{mesh}.

In addition to compatible functionality requirement, the interconnections must be reconfigurable as well, so that spare units can redirect the Input/Output (I/O) channels of the faulty units. Such a reconfiguration is either supported by various built-in mechanisms in configurable systems, such as FPGAs, or by using additional swtiches or Multiplexers (MUXes) in a general way, for Network-on-Chip (NoC) and System-on-Chip (SoC) architectures \cite{banerjee2015conditions} \cite{banerjee2017local} \cite{mesh}. Among the emerging nanoelectronic devices, general reconfiguration capabilities are inherently supported by many device candidates \cite{device1} \cite{device2} \cite{device3}.

If every spare unit can be made available to replace every functional unit, such a maximum flexibility will result in the highest possible reliability. However, achieving such a ``full sharing" architecture is impossible in practice for two main reasons:
First, even though the implementation of spares is becoming progressively cheap as systems scale up, the implementation of the interconnect networks are becoming prohibitively expensive. Not only do the long wires add significant overhead to the area, power and delay, but they also suffer from reliability problems themselves \cite{road} \cite{Beckett} \cite{previous}. Second, for every spare unit to replace every functional unit, the spare units must be functionally compatible with all the functional units in the system. With the exception of homogeneous systems, achieving such a full functional compatibility would be either simply impossible for heterogeneous systems or prohibitively expensive for reconfigurable ones. Hence, spare sharing is limited by strict interconnection constraints and limited functionality match.

Traditianly, many studies of repair-based schemes aimed for achieving the \emph{optimal} fault tolerance scheme; thus, they are usually limited to a very small range of highly simple or regular system structures. Furthermore, the interconnection constraints are not considered as a general limit, but rather as a specific topology such as mesh, ring, or tree structures \cite{graph} \cite{grammars} \cite{mesh}. 

More recent approaches aim for designing repair-based approaches for any spare sharing topology by offering various forms of repair mechanisms:

\begin{itemize}

\item  \textit{Global} spare replacement schemes propose a reassignment of all the spares to all the faulty functional units after each fault occurrence \cite{previous} \cite{banerjee2017local}. Even though such a global mapping approach can in fact take advantage of the entire repair capabilities in the system, it would require a centralized mechanism that comes with significant communication/performance overheads. 

\item \textit{Chained} spare replacement schemes propose a chain of replacements, in which a faulty functional unit is replaced by a chain of other functional units until ultimately replaced by a spare \cite{banerjee2015conditions} \cite{banerjee2017local}. These approaches make it possible to extend the replacement capability of local spares towards far-away functional units globally. However, they usually entail an additional software overhead in the form of rescheduling the tasks among the new set of working units at the end of each repair process, thus resulting in a significant performance overhead for systems with high fault rates.

\end{itemize}

\begin{figure}[tb]
	\centering
	\includegraphics[scale=.42]{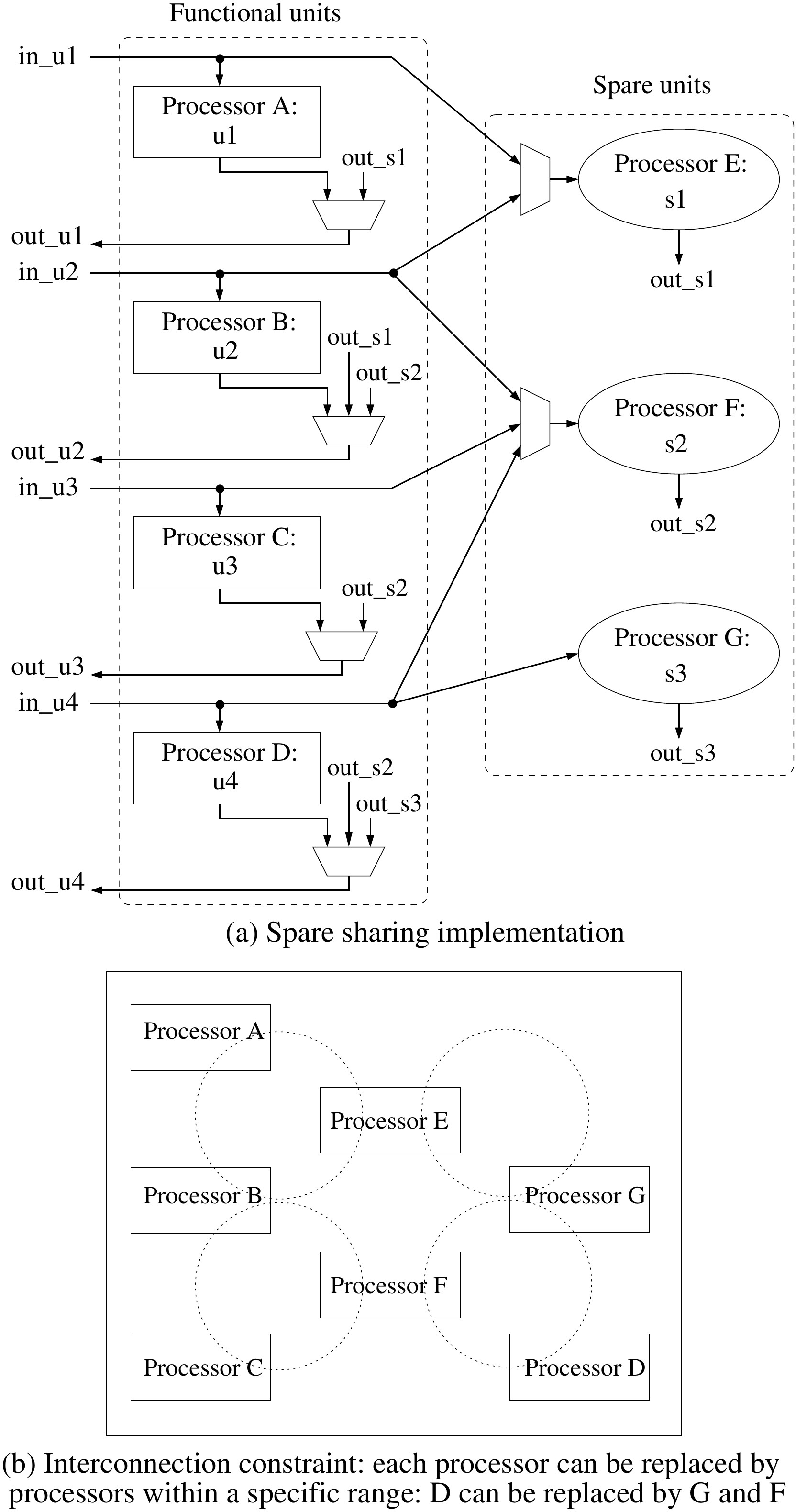}
	\caption{Example of a system with three standby spare processors (E,F,G) shared among four functioning ones (A,B,C,D) in a limited way, due to the interconnection constraints.}
	\label{fig:circuit}
\end{figure}

\begin{figure*}[tb]
	\centering
	\includegraphics[scale=.53]{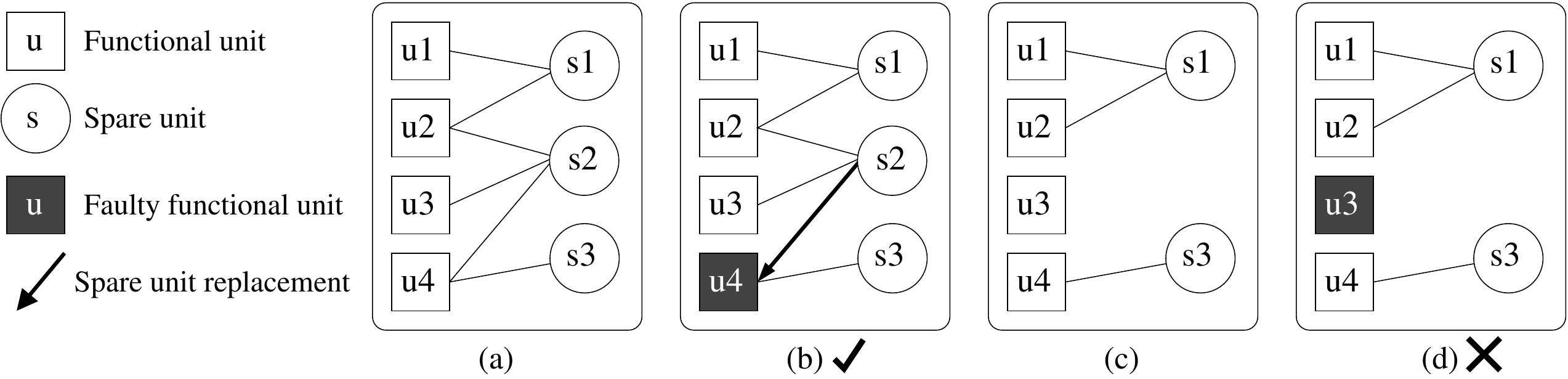}
	\caption{Examples of bipartite graph representation and repair process: (a) the spare sharing network in Figure \ref{fig:circuit}. (b) a successful repair after one fault. (c) the remaining system after the successful repair. (d) system failure: unrepairable after the second fault happening at $u3$.}
	\label{fig:bipartite}
\end{figure*} 

Due to their significant overheads, we argue that such global/chained spare assignment approaches must be used only as a last resort of repair in a spare sharing topology. Therefore, having a fast, low-cost repair mechanism that can be applied to any spare sharing topology of sparse interconnection with high fault rates is the goal of our study. Consequently, this paper proposes an \textit{immediate} spare replacement mechanism, in which only one repair takes place at a time, i.e., one immediate replacement of a functional unit with a spare unit after each fault. As opposed to the global/chained schemes, such an immediate replacement approach can be done locally with minimum disturbance to other functional units, and therefore entails a minimal performance overhead. 

Thus, this paper views the system in the status of immediate-repairable, and tries to maintain such repairability status while performing these low-cost, fast repairs. As a result, the more expensive global/chained approaches will not be invoked every time a functional unit fails, but only when the proposed immediate replacement mechanism fails. After that, a global/chained repair can put back the system in the status of immediate-repairable. Therefore, systems with high fault rates can be mostly repaired at a small overhead using the proposed immediate spare replacement mechanism, and the more expensive, yet optimal, repair approaches will be used only as a last resort of repair.

In this paper, we present two approaches for boosting immediate-repairability of systems with limited spare sharing constraints. First, we present several repair algorithms that select a spare (among a set of candidates) to replace a faulty unit and compare them to identify the one that minimally affects the repairability of the resultant system. 
Second, we propose a design methodology to enhance the repairability of any system with limited spare sharing capability. The proposed enhancement approach works by identifying a set of criteria to pinpoint the ``most vulnerable" functional units and the ``most exploitable" spare units in a network, such that by extending the replacement capabilities of the identified spare units to the identified functional units, maximum gain of repairability boost can be achieved for the entire system. 

The rest of this paper is organized as follows: section II motivates the research work and presents the preliminaries on repairability models; in section III, the proposed repair algorithm is explained; in section IV, the proposed network enhancement methodology for boosting repairability is presented; section V shows the simulation results and section VI concludes the paper.

%--------------------------------------------------------------------------------------------------------------
%--------------------------------------------------------------------------------------------------------------
%--------------------------------------------------------------------------------------------------------------
%--------------------------------------------------------------------------------------------------------------
%--------------------------------------------------------------------------------------------------------------
%--------------------------------------------------------------------------------------------------------------
%--------------------------------------------------------------------------------------------------------------

\section{Motivations and preliminaries}
\label{mot and pre}

We denote the original components in a system as \emph{functional units}, and the standby ones, which are capable of replacing faulty functional units, as \emph{spare units}.

Figure \ref{fig:circuit}(a) shows an example of a Multi-Processor System with four functional units ($u1,u2,u3,u4$) and three built-in spare units ($s1,s2,s3$). Based on the limited sharing network topology, perhaps imposed by locality constraints, shown in Figure \ref{fig:circuit}(b), each spare unit can replace some (but not all) of the functional units. For example, if a fault occurs at $u2$ ($B$), it can be replaced by either spare unit \mbox{$s1$ ($E$)} or $s2$ ($F$), 
yet it cannot be replaced by $s3$ ($G$) or any of the functional units. In order to replace a faulty functional unit with a spare unit, the spare needs to cover the operation of the faulty functional unit and redirect the I/O of the functional unit. This is facilitated by the MUXes, as is shown in Figure \ref{fig:circuit}(a). It must be noted that the overhead (in terms of MUX cost and interconnection cost) increases rapidly as the number of interconnects in the sharing topology increases.

\subsection{System Model}

For a system consisting of a functional unit set, a spare unit set, and the corresponding replacement relationships, it can be uniquely represented by a \emph {bipartite graph} $BG(N_U , N_S , E)$, where $N_U$ and $N_S$ represent the functional unit set and the spare unit set, respectively. Each edge $e=(u, s)\in E$, where $u\in N_U$ and $s\in N_S$, indicates that the corresponding functional unit $u$ can be replaced by the spare unit $s$. 
Figure \ref{fig:bipartite}(a) shows the bipartite graph representation of the example in Figure \ref{fig:circuit}, with squares and circles illustrating functional units and spare units, respectively. 

From the bipartite graph model, it is clear that the amount of redundancy embedded in the system is depicted by $|N_S|$ (number of spare units). The limited spare sharing of the system, imposed by interconnection and functional compatibility constraints, is represented by the edges of the bipartite graph. Specifically, the fan-out degree of a spare unit $s$ (denoted by $d(s)$) illustrates that $s$ can replace any one of the $d(s)$ connected functional units; the fan-out degree of a functional unit $u$ (denoted by $d(u)$) represents the number of accessible spare units for $u$. The overall number of edges in the bipartite graph, $|E|$, approximately depicts the interconnection complexity and implementation cost of the system.

It must be noted that the system in Figure \ref{fig:circuit}(a) is just one example that can be represented by the proposed bipartite graph model. As a matter of fact, such a model can capture the repair behavior of any spare sharing system in which the spare units can take over the jobs of faulty functional units immediately, regardless of the used technology in the system, the types of the units, or the types of the interconnections among the units.

\subsection{Repair Model: Immediate Spare Replacement}

The immediate spare replacement mechanism, in which a faulty functional unit is replaced with an accessible spare unit is represented as follows: after the repair process, the allocated spare unit will be removed with all of its associated edges, and the faulty functional unit node can be marked as fault-free again. 
Figures \ref{fig:bipartite}(b) and  \ref{fig:bipartite}(c) illustrate a repair process. The immediate spare replacement in Figure \ref{fig:bipartite}(b) shows a \textit{successful repair}, in which faulty $u4$ is replaced by $s3$, an the resultant system after repair is shown in Figure \ref{fig:bipartite}(c). Figure \ref{fig:bipartite}(d) depicts an unrepairable fault sequence: no spare unit is available for $u3$ after the second fault, and thus no successful repair exists for this fault. In general, when a faulty functional unit has no accessible spare units, we denote this as a \emph{immediate replacement failure}.

Assuming that a \textit{system failure} occurs when there exists a faulty functional unit that cannot be replaced using any spare units in the network, it must be noted that an immediate replacement failure does not necessarily cause a system failure. Upon an immediate replacement failure, the system might still be repairable via a global spare replacement approach, i.e., a complete reassignment of the already used spare units as well as the unused ones to the faulty functional units, 
For example, upon the immediate replacement failure shown in Figure \ref{fig:bipartite}(d), the system can be repaired by releasing $s2$ from $u4$ and assigning it to faulty $u3$, and using $s3$ to replace $u4$. 

As it was discussed in Section \ref{intro}, by employing efficient immediate spare replacement algorithms and proposing ways to strengthen the topology of spare sharing networks, this work tries to maintain the repairability of the system for the immediate spare replacement mechanism. This way, the more expensive approaches of global/chained spare replacements will be invoked less frequently, only upon an immediate replacement failure of the proposed model.

\subsection{Repairability}

\begin{figure}[tb]
	\centering
	\includegraphics[scale=.51]{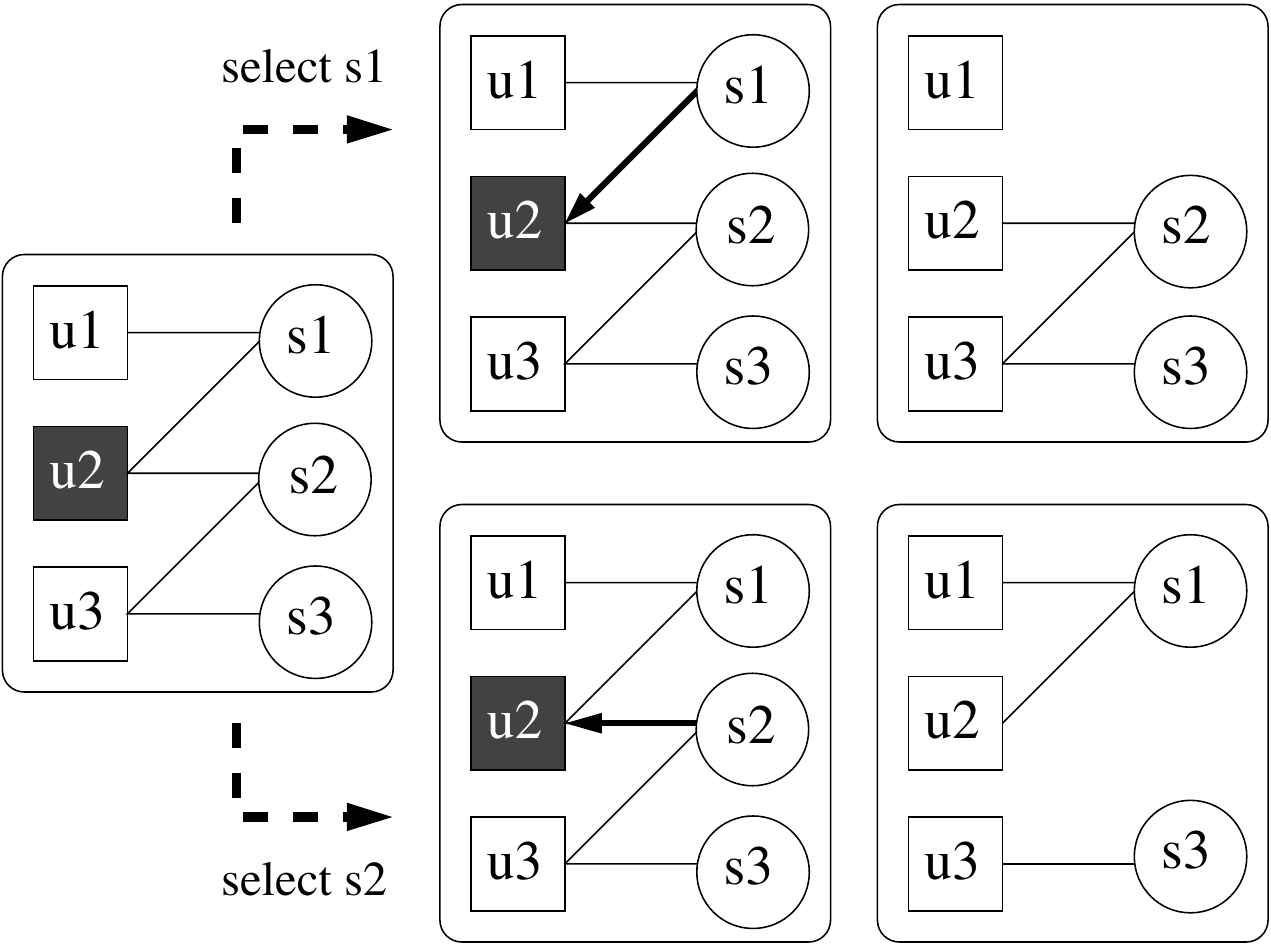}
	\caption{Example of different resultant systems according to distinct spare unit selections: the repair process performed at bottom guarantees the remaining system to survive the next fault, but the one at top might fail if the next fault happens at $u1$.}
	\label{fig:repair}
\end{figure}

The repairability of a system (under the immediate spare replacement mechanism) is determined jointly by three major factors:

\begin{enumerate}
\item \emph{Fault Sequence}: A immediate replacement failure may or may not occur depending on the specific sequence of fault occurrences. For example, in Figure \ref{fig:bipartite}(c), a fault occurring at $u3$ cannot be repaired (shown in Figure \ref{fig:bipartite}(d)); however, if a fault occurs at $u1$ or $u2$, the system can be repaired by $s1$.
\item \emph{Network Topology}: The repairability of a system heavily depends on the amount of redundancy (number of spare units) and the topology of spare sharing of the bipartite graph network.

\item \emph{Replacement Algorithm}: Each functional unit can be replaced by any of its connected spare units upon fault occurrences. Thus, whether or how long a system can survive, depends on the spare unit selection made upon every fault occurrence. Figure \ref{fig:repair} shows an example of two different resultant systems for the same fault, according to two distinct selections made for spare unit.
\end{enumerate}

To determine which of the resultant systems in \mbox{Figure \ref{fig:repair}} is more repairable, we adopt the model of  \cite{DFTpaper} \cite{previous}. Repairability is measured by the survival probability of a system, for all the possible fault occurrence sequences with repair options. With this repairability model, we assume: \mbox{1) independent} fault occurrence on each functional unit (and spare unit if the spare is in use) with equal probability, and 2) sequential occurrences of faults, i.e., one repair is performed after every single fault occurrence. For $f$ faults occurring sequentially on the $|N_U|$ functional units, there is a total number of $|N_U|^f$ equally possible fault occurrence patterns, including repeated fault occurrence onto the same functional unit (i.e., the spares that take over its job). For any of the $|N_U|^f$ fault occurrence patterns, it is defined as repairable if at least one successful repair sequence exists for it \cite{DFTpaper} \cite{previous}.

\emph{\textbf{Definition}}: The repairability of a system $BG(N_U , N_S , E)$ under $f$ faults, $RE(BG,f)$, is defined by the percentage of cases that can be repaired (using an immediate spare replacement) over all the $|N_U|^f$ possible fault occurrence patterns. 

Plotting the entire repairability curve $RE(BG,f)$ for a given network is generally impractical, due to the huge number of possible fault sequences. However, it is possible to deduce a number of characteristics of the repairability curve \cite{previous}, also shown in Figure \ref{fig:curve}(a):

\begin{figure}[tb]
	\centering
	\includegraphics[scale=.34]{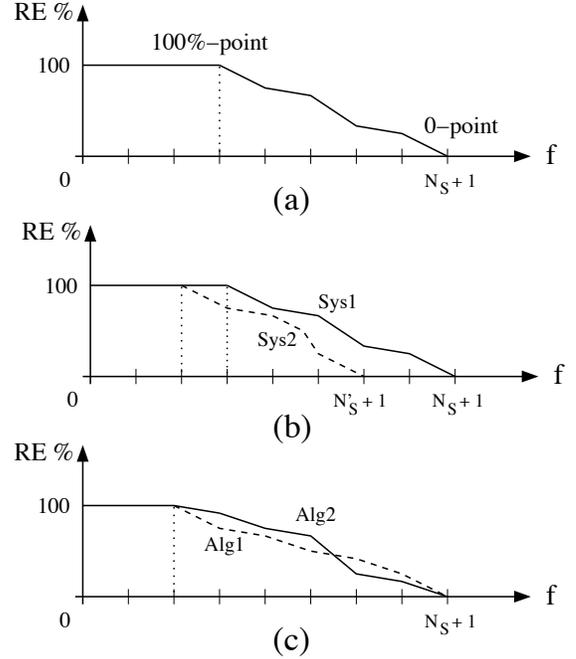}
	\caption{(a): repairability plot of a general system with its main points. (b): repairability plots of two different systems. (c): repairability plots of a same system for two different algorithms}
	\label{fig:curve}
\end{figure}

\begin{enumerate}
\item \emph{Monotonically decreasing}: the repairability of a system decreases monotonically as the number of fault increases \cite{previous}. 
\item \emph{100\%-point}: a system can always survive the next fault as long as all functional units have at least one connected spare unit. Consequently, if the number of fault occurrences is less than the minimum fan-out degree in the functional unit set, repairability remains at 100\%. 
\item \emph{0-point}: because each faulty functional unit is replaced by a spare unit, the maximum number of faults that a system can tolerate is no more than the number of spare units, i.e., repairability function always drops to 0 (guaranteed immediate replacement failure as well as complete system failure) when the number of fault occurrences is greater than $|N_S|$.
\end{enumerate}

\subsection{Repairability Boost}
The repairability of a system can be enhanced in two ways:
\begin{enumerate}

\item \emph{Replacement Algorithm} targets the repair process. The repairability of a system can be enhanced based on a carefully designed spare unit selection algorithm, upon each fault occurrence. Various spare unit selection algorithms are suggested in \cite{mesh} \cite{graph} \cite{previous}. Figure \ref{fig:curve}(c) depicts the repairability curves of a same system, obtained from two different algorithms. As it can be seen from the figure, the 100\%-point and 0-point are the same for both algorithms (as they are determined by the topology of the initial bipartite graph only); however, the two algorithms result in different repairability curves between these two points: Alg2 outperforms Alg1 for the first few faults happening after 100\%-point, while Alg1 outperforms Alg2 towards the final faults.

\item \emph {Network Construction} concerns the topology of the bipartite graph, i.e., the number of functional units, spare units and the sharing structure. The repairability of a system heavily depends on its network topology. Figure \ref{fig:curve}(b) shows the repairability curves of two different systems: The overall trend of the repairability curve including the position of the 100\%-point and 0-point are different for these systems. In this example, Sys1 is obviously more repairable than Sys2.

\end{enumerate}

An important question to ask is that given two repairability curves, which one is the better one? The answer to this question for the repairability curves of Figure \ref{fig:curve}(b) is rather easy, as the repairability of Sys1 is consistently higher than that of Sys2. But for the given curves in Figure \ref{fig:curve}(c), the answer would depend on the application of the problem. The area under curve might be a good metric to show the overall repairability of a system; however, it may be the case that it is preferred for the repairability to be as high as possible for the first few faults. If that is the case, then Alg2 is a better choice. In other words, different algorithms may tend to result in different repairability curves; and it will be up to the designer to select the best algorithm based on its requirements.

In this paper, we will focus on both of these approaches. First, we propose several spare selection algorithms and discuss the performance of each algorithm. Second, we present a methodology of boosting system repairability based on the limited spare sharing constraints. To enhance the repairability of any given spare sharing network, the proposed approach works by: \mbox{1) identifying} a set of criteria to pinpoint the ``most vulnerable" functional units and the ``most exploitable" spare units in a network; and 2) adding a few extra connections between the identified units, to achieve the maximum gain of repairability boost for the entire system.

%--------------------------------------------------------------------------------------------------------------
%--------------------------------------------------------------------------------------------------------------
%--------------------------------------------------------------------------------------------------------------
%--------------------------------------------------------------------------------------------------------------
%--------------------------------------------------------------------------------------------------------------
%--------------------------------------------------------------------------------------------------------------
%--------------------------------------------------------------------------------------------------------------

\section{Replacement Algorithm}

This section targets the process of repairing the system due to the faults happening during run-time. Basically, the following question is to be asked: if a functional unit fails in a given bipartite graph, which of the candidate spare units (those connected to the faulty functional unit) must replace it, such that system repairability is affected minimally?  In the extreme cases of a ``fully connected" network (where all the spares can repair all the functional units) and a ``dedicated" network (where the functional units do not share any of their spares), the replacement algorithm does not come into play, because no matter which of the spare units replaces a faulty functional unit, the resultant network would be the same. However, for a network with limited spare sharing, replacemedt algorithms play an important role towards the system repairability \cite{previous}.
 
\subsection{Motivation}
\label {mot-algorithm}
When a spare unit is selected to replace a faulty functional unit, the group of other functional units that are originally connected to this spare unit will have one less supportive spare unit, as it cannot be used to replace those any more. To select the spare unit with the least side effect of this kind, an already widely ``popular" spare unit (with many connected functional units) should not be picked, because a wide neighborhood of functional units will then be affected. 

However, popularity is not the sole criterion for selecting a spare unit. If some of the originally connected functional units to a spare unit do not have much access to other spare units, then, regardless of being popular or unpopular, such an ``essential" spare unit (with vulnerable functional unit(s) connected to it) should not be picked, because otherwise such vulnerable functional unit(s) would have even a reduced chance of support from the spare unit; thus affecting the repairability of the entire system, as it heavily depends on protecting the vulnerable functional units.

\subsection{Terminology}
\label{pop-urg}

In this section, we formally define the above-mentioned properties for functional units and spare units.

In a bipartite graph $BG(N_U,N_S, E)$, the adjacent node set $adj(s)$ for spare unit $s$ denotes the set of functional units connected to the node \mbox{$s: adj(s) = \{u :(u, s)\in E\}$}. 
Let $MinDeg(N)$ denote the minimum degree among a node set \mbox{$N$: $MinDeg(N) = min \{d(x), \forall x\in N\}$}. 
For a spare unit $s$, then, $MinDeg(adj(s))$ denotes, for the least supported functional unit in the neighborhood of $s$, the number of its accessible spare units. Also, let $d(.)$ be used to denote the fan-out degree of the functional units and the spare units in a network. Throughout the paper, the \textit{neighborhood} of a functional unit refers to the spare units that are connected to it, and vice versa.

The following three properties (one defined for functional units and two defined for spare units) will be used in the proposed schemes.

\begin{itemize}
\item \textit{Vulnerability}: 
The vulnerability of a functional unit $u$ is denoted by its fan-out degree $d(u)$, i.e., how many spare units can replace it. Consequently, $u1$ is weaker (more vulnerable) than $u2$ iff $d(u1) < d(u2)$. Similarly, $u2$ is stronger (less vulnerable) than $u1$.

\item \emph{Popularity}:
The popularity of a spare unit $s$ is denoted by its fan-out degree $d(s)$, i.e, how many functional units it can replace. Consequently, $s2$ is more popular than $s1$ iff $d(s2) >  d(s1)$.

\item \emph{Essentiality}:
The essentiality of a spare unit $s$ is defined as the fan-out degree of the most vulnerable functional unit in its neighborhood, and is denoted by $MinDeg(adj(s))$. 

This criterion can show whether a spare unit is essential for some functional units. In other words, if $MinDeg(adj(s1))$ is large for some spare unit $s1$, it shows that even the weakest functional unit in the neighborhood of $s1$ has a lot of other accessible spare units. On the other hand, if $MinDeg(adj(s2))$ is small for some spare unit $s2$, it indicates that at least one of the functional units in its neighborhood has very limited access to other spare units other than $s2$, thus is relying heavily on $s2$ to survive. Consequently, $s2$ is more essential than $s1$ iff $MinDeg(adj(s2)) <  MinDeg(adj(s1))$. 
\end{itemize}

The motivation for defining the essentiality property as stated above is that if a spare unit serves only strong functional units (with accessibility to many other spare units), then the essentiality for this particular spare unit is low, and the side effect of picking this non-essential spare unit is also low. Since the repairability of the system heavily depends on the most vulnerable functional units in the network, such essentiality of a spare unit can be captured by the minimum degree among the functional units connected to this spare unit.

\subsection{PE Algorithm: Preserving Essential spare units}
\label{PE}
In the worst-case scenario, in which all the faults happen at the most vulnerable functional unit (with the minimum fan-out degree), no matter what algorithm is used, the whole system fails after $MinDeg(N_U)$ faults, as there would be no more spare unit to replace it.
In another case, if the first fault happens at a functional unit that does not share any spare unit with the most vulnerable functional unit, then the system is guaranteed to survive at least $MinDeg(N_U) + 1$ faults (in the worst-case scenario when all the next faults happen at the most vulnerable functional unit), no matter what algorithm is used. 
However, if the first fault happens at a functional unit that shares a spare unit with the most vulnerable functional unit, then selecting the spare unit becomes important. If the \textit{shared} spare unit is selected to replace the faulty unit, then the system is only guaranteed to survive $MinDeg(N_U)$ faults, as the repair choice for the first fault reduces the fan-out degree of the most vulnerable functional unit by 1, and the next faults could all happen at the same vulnerable functional unit in the worst-case. However, if any other spare unit (not the shared one) is selected to replace the faulty functional unit, the system is guaranteed to survive $MinDeg(N_U) + 1$  faults in the worst-case scenario. This argument could be extended to form an algorithm, suitable to achieve maximum survival of a system under the worst-case scenario.

Generally, each spare unit is shared among several functional units. Every time a spare unit is used to replace a faulty functional unit, all of its adjacent functional units will have one less edge. This loss of one edge is very crucial for the most vulnerable functional unit in the neighborhood of the selected spare unit, as it could determine the overall repairability of the system. When a functional unit fails, one of its adjacent spare units must be selected to replace it. Each of these candidate spare units has its own most vulnerable functional unit in its neighborhood. The overall survival of the system heavily depends on these most vulnerable functional units. In order to guarantee the maximum survival of the system under the worst-case scenario for the upcoming faults, one must select the least essential spare unit, whose most vulnerable functional unit is the strongest among other spares' most vulnerable functional units.

\noindent
\textbf{Preserving Essential spare units (PE)}: If functional unit $u$ fails, among all of its adjacent spare units in $adj(u)$, the least essential spare unit must replace it, i.e., the one whose most vulnerable functional unit in its neighborhood is the strongest among other spares' most vulnerable functional units.  More formally, for a spare unit $s$, then, $MinDeg(adj(s))$ denotes, for the least supported functional unit in the neighborhood of $s$, the number of its accessible spare units. 
Therefore, the spare unit with $max[MinDeg(adj(s))]$, where $s \in adj(u)$, will replace the faulty functional unit.

By preserving the essential spare units, the PE algorithm guarantees to maintain the maximum repairability of the system under the worst-case scenario by ensuring that vulnerable functional units will have their maximum support for the upcoming faults.

\subsection{PP Algorithm: Preserving Popular spare units}

According to the previous algorithm (PE), a widely popular spare unit may be selected to replace a faulty unit if it has a strong neighborhood of functional units. However, this could impact the overall repairability of the system in the \textit{long-future}, as a widely popular spare unit is capable of replacing many functional units. Our second algorithm is based on the observation that a widely popular spare unit can be used to replace many functional units. Therefore, it must be preserved over less popular spare units, i.e., replacing a faulty functional unit with a widely popular spare unit will have a side-effect on a wide range of functional units, whereas selecting a less popular spare unit only affects a small number of functional units.

\noindent
\textbf{Preserving Popular Spares (PP):} If functional unit $u$ fails, the least popular spare unit in its neighborhood must replace it. More formally, the spare unit with $min[d(s)]$, where $s \in adj(u)$, will replace the faulty functional unit. In other words, a spare unit, for which $d(s) = MinDeg(adj(u))$ must be selected to replace $u$.

By preserving the popular spares, the PP algorithm aims at achieving high system repairability for the faults happening towards the long-future; this is due to the fact that those spares with large fan-out degrees will remain in the network, in order to serve many functional units towards the end.

\subsection{Combined Algorithms}

In a sparse network, it often occurs that there exist many functional/spare units tied with the same minimum fan-out degree. Even for a network with a lot of edges, the system will become sparse towards the final faults. In such cases, which on average contributes to 21\% of the decisions in our case-study networks, both PE and PP algorithms require a tie-breaker. Since each of these algorithms considers an important aspect of a network, each can be used as a tie-breaker for the other scheme. Therefore, we study two combined algorithms:

\noindent
\textbf{PE+PP:} using PE as the main algorithm and PP as the tie-breaker.

\noindent
\textbf{PP+PE:} using PP as the main algorithm and PE as the tie-breaker.

It must be noted that even though it is possible to propose more complicated and efficient replacement algorithms, they may not be easily implementable in a local decentralized approach, i.e., they may need either a centralized scheme to or a lot of decentralized local communications (both resulting in high performance overheads) before making the final replacement decision. the advantage of the above-mentioned algorithms is that they can be easily implemented in a decentralized fashion with very small performance overhead.

\begin{figure*}[tb]
	\centering
	\includegraphics[scale=.38]{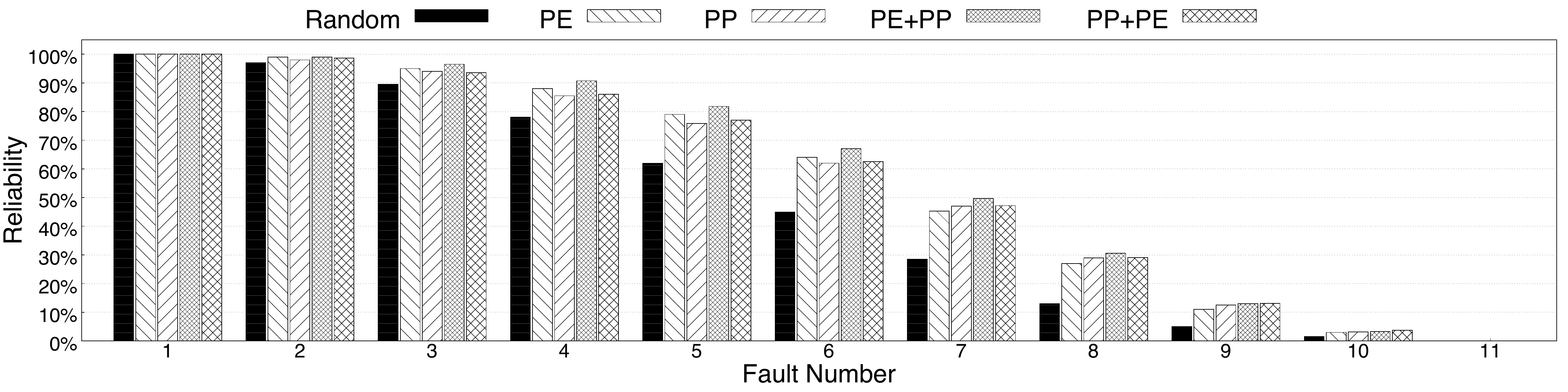}
	\caption{Comparison of different repair algorithms on systems with 15 functional units, 10 spare units, and 40 edges.}
	\label{fig:algs}
\end{figure*}

\subsection{Simulation Results}
In this section, we present the evaluation of the proposed repair algorithms. To do so, we implemented the algorithms on a wide range of networks (the functional units ranging from 10 to 50, the spare units ranging from 5 to 30, and edges from 20\% to 60\% of the network capacity). As the results were mostly consistent across the experiments, we show the evaluation of the algorithms on networks with 15 functional units, 10 spare units, and 40 edges that are randomly placed between the functional units and the spare units. All results are the averages over 100 runs of network generation. The repairability of every system is evaluated by the percentage of repairable sequences over 10,000 randomly generated fault sequence patterns. 

Figure \ref{fig:algs} shows the system repairability for 5 different algorithms: Random, PE, PP, PE+PP and PP+PE. For PE and PP algorithms, the random selection of spare units is used as a tie-breaker. As it can be seen from the graph, all schemes significantly outperform the random allocation. Figure \ref{fig:algs} depicts the applicability of PE and PP algorithms. As we discussed before, PE has a better performance for the near future (faults up to the 6th one in this example), as it preserves the vulnerable functional units, while PP works better towards the last faults (from the 7th to the 10th), because more popular spare units are preserved for the last faults, so more functional units have access to some spare units. 

The figure also depicts the importance of choosing a tie-breaker: PE+PP outperforms all other algorithms for almost all the faults (except for the last two faults, where the results are comparable between all algorithms, as the repairability is very low at this point). The high performance of PE+PP is due to the fact that a lot of tie situations happen during the repair process. PE makes sure to maintain the repairability of the system for near future, and in case of a tie, PP helps to preserve the more popular spares towards the end faults. An interesting observation is that PP+PE does not outperform PP. This is because the chances of having spare units being tied in maximum degree is relatively low, due to the small number of spare units involved in the selection process for each fault (as opposed to the chance of having poor functional units in a tie situation, due to a large number of functional units involved in the selection process). Therefore, most of the tie situations happens towards the end, when the system is sparse; however, PE would not be a good choice at this point, because most of the functional units are almost equally vulnerable.

Overall, the PE+PP algorithm shows the best performance, as it considers the urgent need of preserving vulnerable functional units first and preserves the popular spares for later faults if possible. Furthermore, based on the theoretical argument in section \ref{PE}, for a given system, PE can guarantee the maximum system survival (in terms of the number of faults the system can tolerate), under the worst-case scenario, which makes it a good choice for many applications, where all functional units must be working for a system to run.

%--------------------------------------------------------------------------------------------------------------
%--------------------------------------------------------------------------------------------------------------
%--------------------------------------------------------------------------------------------------------------
%--------------------------------------------------------------------------------------------------------------
%--------------------------------------------------------------------------------------------------------------
%--------------------------------------------------------------------------------------------------------------
%--------------------------------------------------------------------------------------------------------------

\section{Network Construction Enhancement}
This section targets the process of constructing a spare sharing network for enhancing system repairability.
According to the repairability model, an immediate replacement failure happens when a faulty functional unit cannot be replaced by any spares, i.e., the fan-out degree of the faulty functional unit is equal to zero. Consequently, the best way to construct a network would be to distribute all edges evenly in a regular manner among functional units and spare units \cite{DFTpaper} \cite{previous}. A high-order \emph{ring} structure, for which all the functional units have a same fan-out degree, and so do all the spare units, can offer such an ``optimal" repairability for a given number of edges. However, due to several interconnection constraints such as wiring area, routing delays, and power consumption, as well as the functionality match constraint, it is not always possible or desirable to have a perfect ring structure network \cite{DFTpaper} \cite{previous}.
In this work, we focus on the general framework of improving the repairability of any arbitrary network, with sparse sharing of spare units. 

The proposed network enhancement technique tries to increase the system repairability up to the maximum extent through a small modification of the network topology in the form of adding a few extra interconnections between existing functional units and spare units. Basically, we focus on this question: given a small budget (in the form of one (or a few) extra connection(s)), how should the extra connections be added, such that system repairability is boosted to the largest extent? Contrary to the replacement algorithms, to properly add the extra connections for boosting network repairability, one should focus on the selection of both spare units and functional units. To do so, we develop a set of criteria to pinpoint: \mbox{1) the most} vulnerable functional unit, and 2) the most exploitable spare unit, of a given network, so that a priority list of extra edges can be provided for most gains in repairability boost.

\subsection{Motivation}
\label {mot}

From the perspective of a functional unit, adding an extra connection has the obvious benefit of expanding access to one more spare unit. The question is then which functional unit within a given network can benefit the most. Naturally, the functional units with the least accessible spare units are more likely to benefit from such extra connectivity. According to the repairability function, the entire repair process fails if any faulty functional unit has no more accessible spare units. Therefore, any extra budget to strengthen one of the functional units should go to the most ``vulnerable" one (with the minimum accessible spare units). 

The effect of adding an extra connection from the perspective of a spare unit is not as clear as it is in the case for a functional unit. On one hand, it makes a spare unit accessible to more functional units, which indicates an enhancement to repairability. On the other hand, supporting a new functional unit by a specific spare unit has a side effect on the group of functional units that are originally connected to this spare unit: this originally supported group of functional units will have reduced chance of support from this spare. Therefore, choosing a widely popular spare unit can affect a large neighborhood of functional units. Accordingly, the side effect of choosing a less popular spare unit would be smaller. However, it may be the case for a less popular spare unit to be essential for some of the functional units in its neighborhood; therefore, selecting an essential spare unit can significantly weaken the most vulnerable functional units of the network, on which, the entire system repairability depends.

\subsection{Spare unit selection}
Since we are not adding new spares, the key idea in selecting a spare unit is to find the ``most exploitable" one among all.

 \begin{figure} [tb]
	\centering
	\includegraphics[scale=.55]{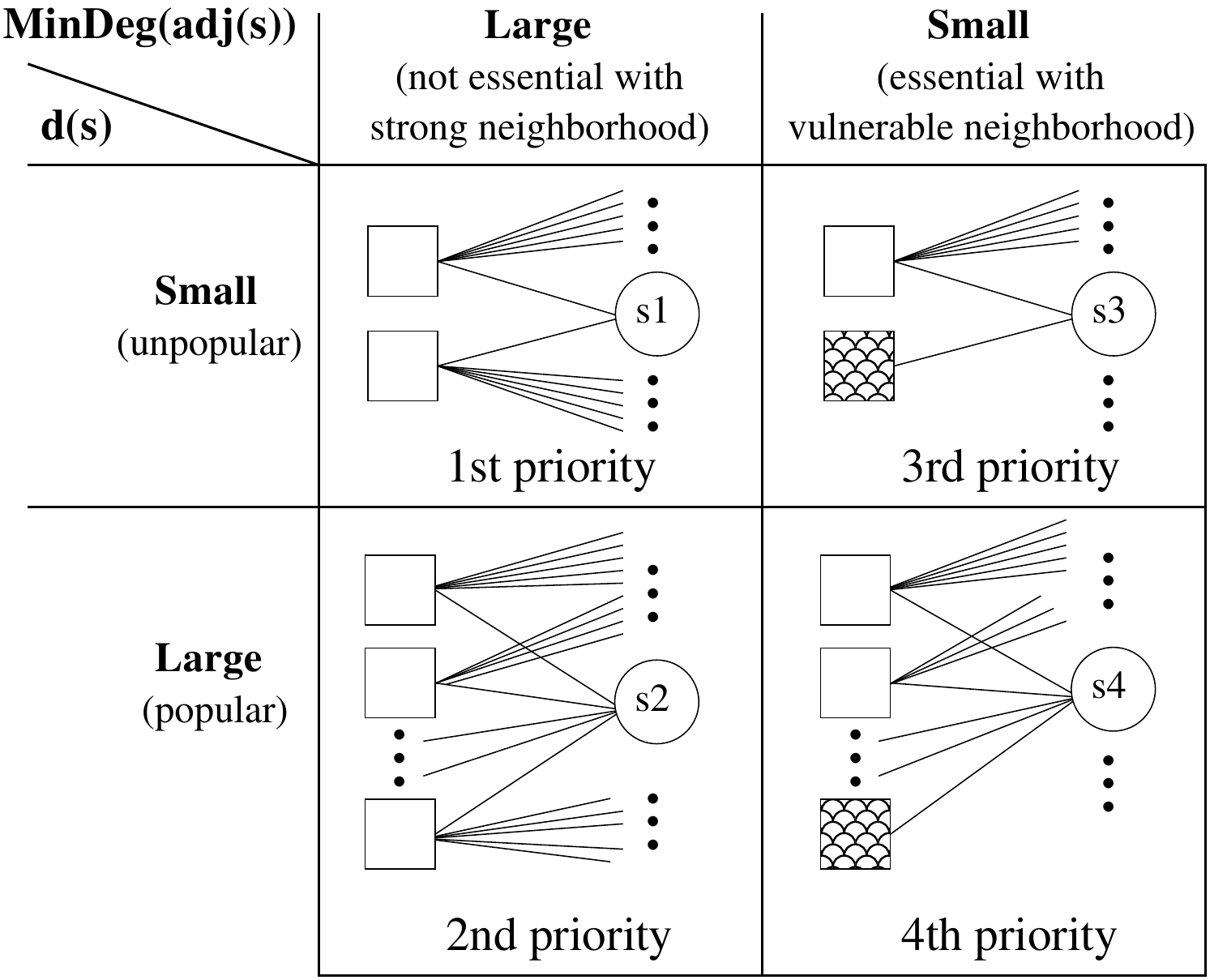}
	\caption{Spare units selection priorities based on popularity and essentiality (functional units filled with the pattern are the most vulnerable ones). }
	\label{fig:spare}
\end{figure}

According to our discussion on the criteria of popularity and essentiality, clearly, the most exploitable spare unit is the one that is neither essential (with a large $MinDeg(adj(s))$) nor popular (with a small $d(s)$). By the same logic, the least priority goes to the spare unit that is both essential and popular. Since system repairability is dominated by the most vulnerable functional unit, the second priority should be to avoid picking an essential spare unit. If $MinDeg(adj(s))$ is large for a popular spare unit, functional units in its neighborhood are strong enough and do not need it urgently. This makes it a good candidate for extending the accessibility to a new (vulnerable) functional unit.
Figure \ref{fig:spare} provides the priority ranking for spare unit selection based on the two criteria, namely the $MinDeg(adj(s))$ and $d(s)$. 

The principles shown in Figure \ref{fig:spare} for selecting a spare unit to receive an extra edge are implemented by: 
\begin{enumerate}
\item ranking all the spare units based on essentiality \mbox{($MinDeg(adj(s))$)} and choose the largest one(s);
\item using popularity ($d(s)$) as a tie-breaker to choose the least popular spare unit among the top ranked ones \mbox{in 1).}
\end{enumerate}

\subsection{Functional unit selection}

\begin{figure}[tb]
	\centering
	\includegraphics[scale=.53]{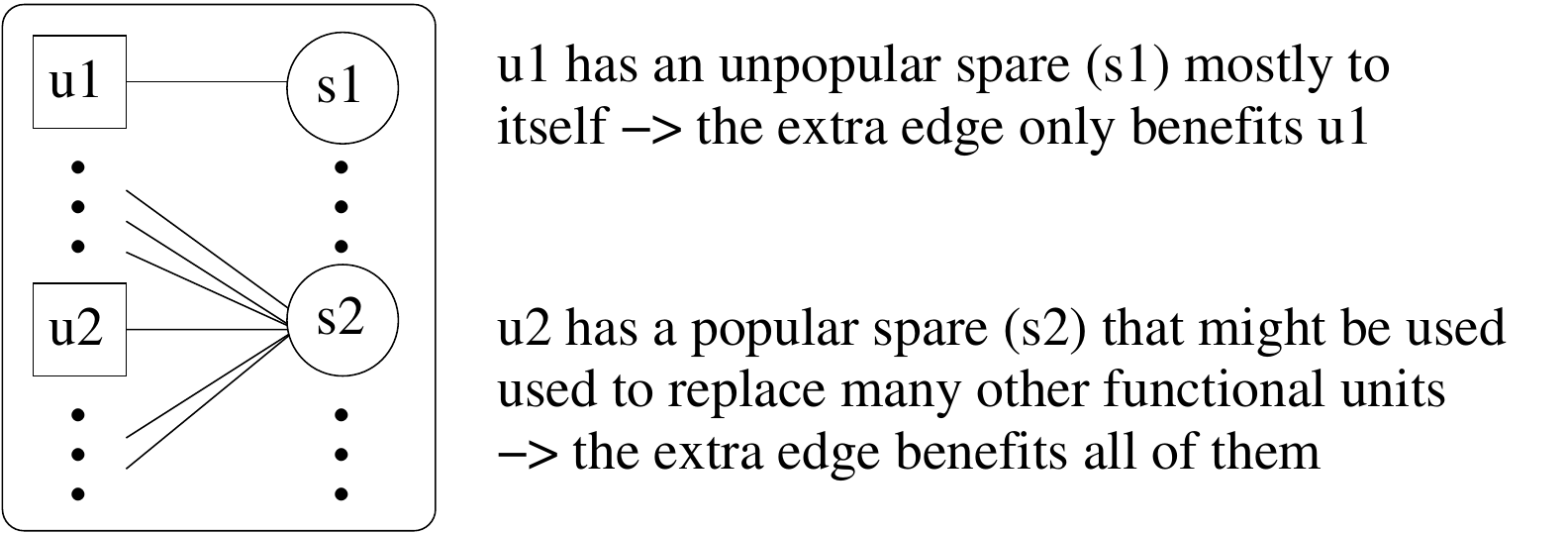}
	\caption{Example of functional unit selection tie-breaker.}
	\label{fig:functional}
\end{figure}

We have motivated in Section \ref{mot} that the most vulnerable functional unit (with the minimum fan-out degree) should be selected for the most gain in repairability boost. 

In a sparse network, it often occurs that there exist many functional units tied with the same minimum fan-out degree. In such cases, a careful observation reveals a secondary criterion, which can be used as an effective tie-breaker. Figure \ref{fig:functional} shows an example of two functional units ($u1$ and $u2$), tied in degree, each with only one accessible spare unit ($s1$ and $s2$ respectively). However, $s1$ is less popular than $s2$. In this case of deciding which between $u1$ and $u2$ should receive the extra ``help", $u2$ is a better choice, because increasing the number of accessible spare units for $u2$ will relieve the urgent demand from $u2$ to $s2$, thus indirectly benefit all the neighboring functional units of $s2$. Such a secondary repercussion of benefit is more significant in the case of $u2$, because $s2$ has a larger neighborhood compared to $s1$. Therefore, it should be selected over $u1$.
In other words, among all the tied functional units ($\forall u| d(u)=MinDeg(N_U)$), the one with the largest $MinDeg(adj(u))$ should be selected. Here, $MinDeg(adj(u))$ essentially captures the popularity of the spare units in the neighborhood of functional unit $u$. 

Therefore, the selection of a functional unit for an extra edge is done by: 
\begin{enumerate}
\item ranking all the functional units based on vulnerability \mbox{($d(u)$)} and choose the smallest one(s);
\item using the functional unit with maximum $MinDeg(adj(u))$ as a tie-breaker to benefit more functional units.
\end{enumerate}

\subsection{Simulation Results}

In this section, we will evaluate the proposed network enhancement scheme, in terms of performance, cost and scalability.
In order to evaluate the repairability gain of the network enhancement scheme, we implement it on a wide range of networks (with different number of functional and spare units and connections among them). As the results were consistent, we present the evaluation for networks with 15 functional units, 10 spare units, and initialized with 20 connections randomly placed between the functional units and the spare units. All results are the averages over 100 runs of network generations. The repairability of every system is evaluated by the percentage of repairable sequences over 10,000 randomly generated fault sequence patterns. We do not impose any specific repair algorithm, i.e., faulty functional units are replaced by one of their accessible spare units selected randomly.

\begin{figure}[tb]
	\centering
	\includegraphics[scale=.49]{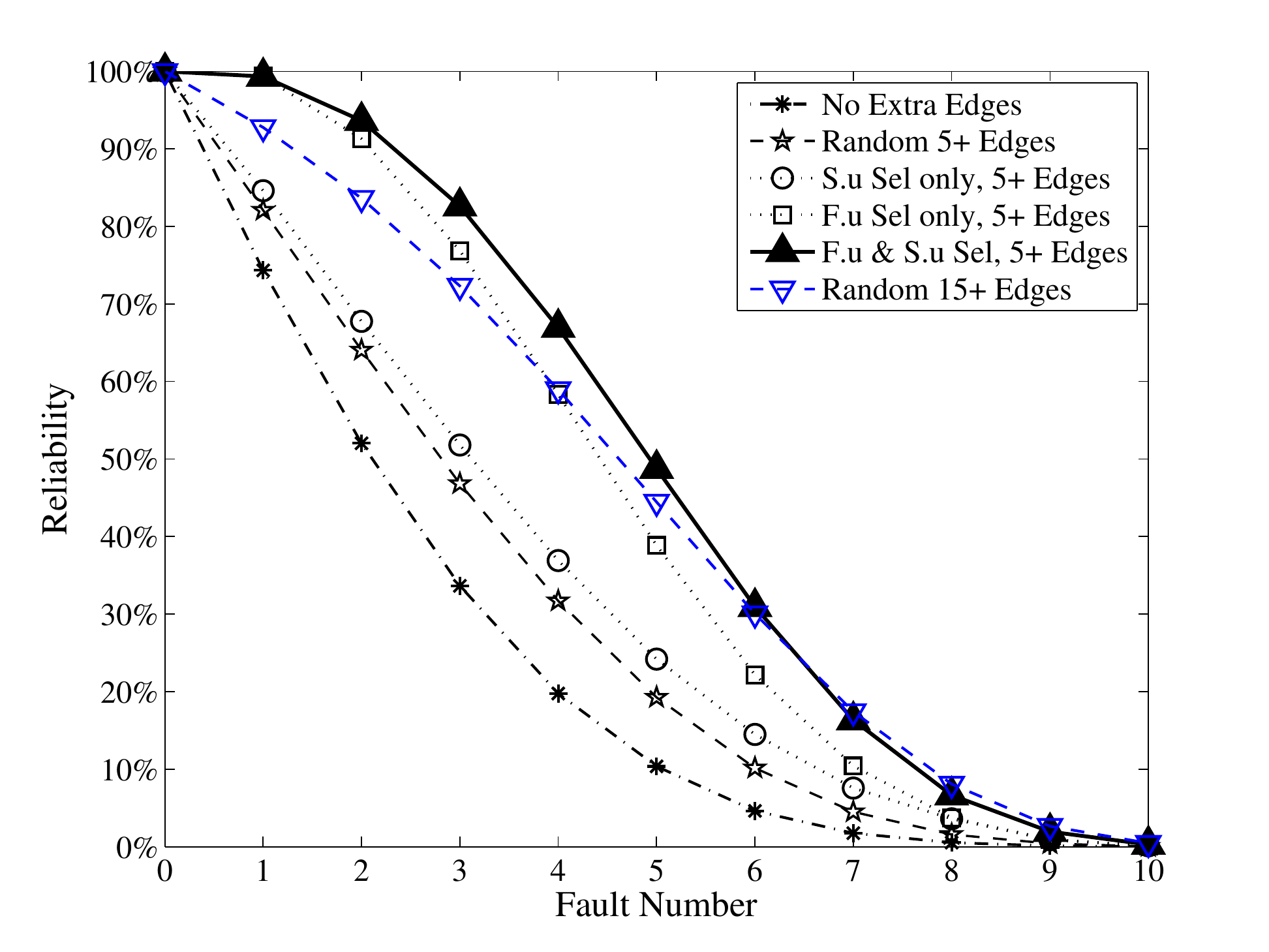}
	\caption{repairability boost evaluation of networks with 15 functional units, 10 spare units, and 20 initial edges.}
	\label{fig:result_1_eval_cost}
\end{figure}

Figure \ref{fig:result_1_eval_cost} shows the repairability of the original network (with no extra edges) and the repairability after adding 5 extra edges to the original network using various methods: 1) spare units and functional units are both selected randomly, 2) only the most exploitable spare units are selected, to be connected to a random selection of functional units, 3) random spare units are connected to the most vulnerable functional units, and 4) both the most exploitable spare units and the most vulnerable functional units are selected to be connected.

\begin{figure}[tb]
	\centering
	\includegraphics[scale=.49]{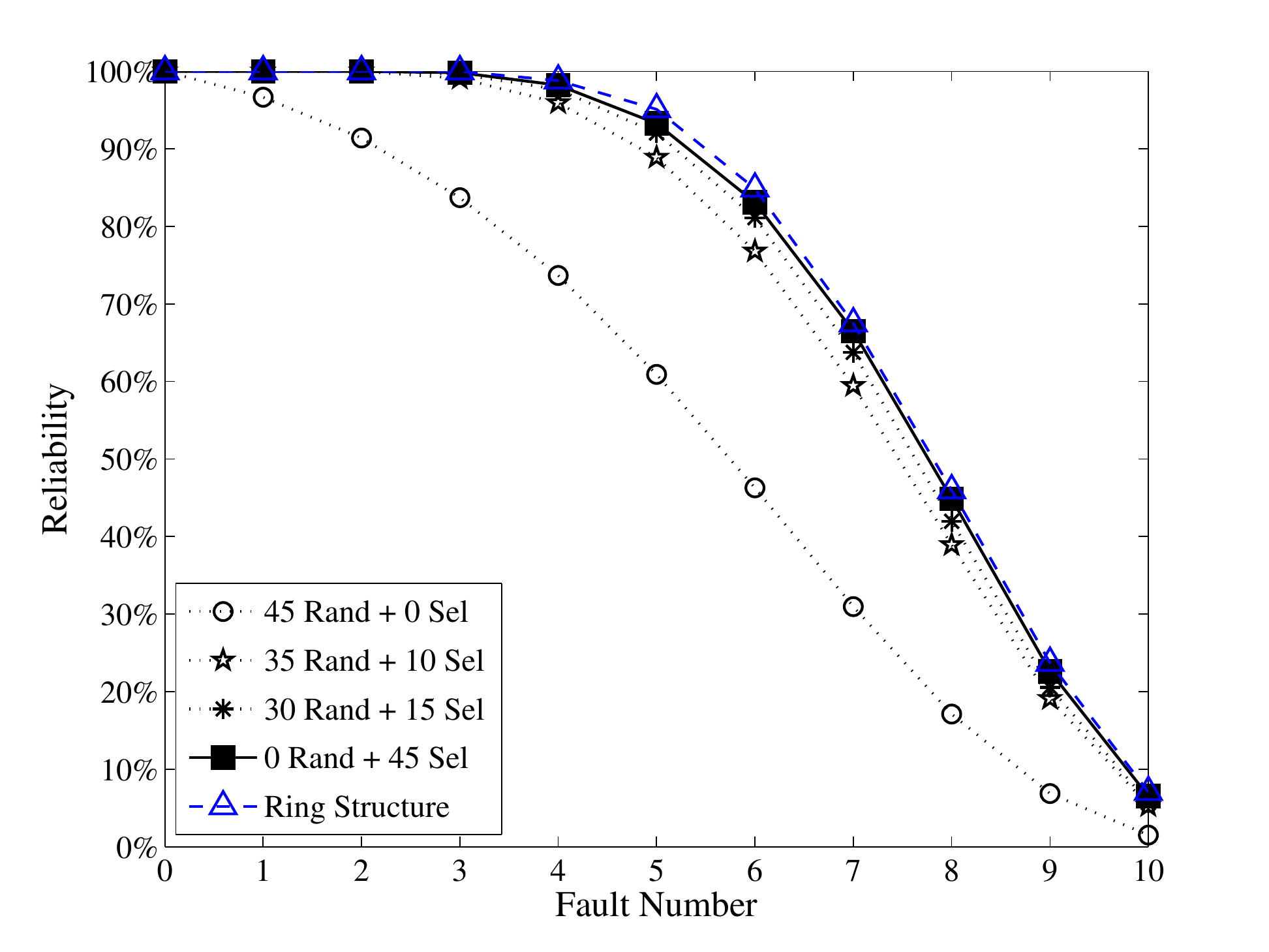}
	\caption{Various approaches for building a network with 15 functional units, 10 spare units, and 45 edges.}
	\label{fig:result3_build}
\end{figure}

As can be seen in Figure \ref{fig:result_1_eval_cost}, even randomly added connections can boost repairability over the original network, especially at the later stage of fault occurrence sequences. 
More importantly, the selection of the extra edges affects repairability boost significantly. Figure \ref{fig:result_1_eval_cost} shows that, selecting the most exploitable spare units (to ``help" a set of random functional units) can improve the average repairability by 48\% compared to the original network  (without extra edges), and 12\% compared to the case of fully randomly selected edges. Selecting the most vulnerable functional units (and random spare units) results in better average repairability improvements by 103\% compared to the original network and 54\% compared to the case of fully randomly selected edges. Apparently, functional unit selection is more crucial than spare unit selection, because preserving the most vulnerable functional unit is vital to avoid system failure. 

The largest repairability boost is obtained by selecting both the most vulnerable functional units and the most exploitable spare units. In this case, the average repairability improvement is 127\% compared to the original network and 70\% compared to the fully random selection. 

Figure \ref{fig:result_1_eval_cost} also illustrates the cost efficiency of the proposed approach. It shows that by carefully selecting the functional units and spare units, only a few extra edges (in this case, 5) are needed to achieve significant repairability gain, which is more than twice (2.27 times) of that in the original network. A random selection-based network enhancement, however, would need 3 times the cost (15 extra edges in this case) to reach comparable repairability results.

Figure \ref{fig:result3_build} depicts a spectrum of approaches for building a network with 15 functional units, 10 spare units and 45 edges. At one end of the spectrum, the network is constructed entirely randomly, while at the other end, the proposed methodology is used to add all the edges, one at a time. Two other networks in the middle are shown by starting with 35 random edges plus adding 10 edges using the proposed methodology, and 30 random edges plus 15 selected edges, respectively. These approaches are compared against a network constructed with the highest repairability, when all the 45 edges are distributed evenly in a regular manner, forming a high-order ring structure. As it can be seen, the most repairability boost is resulted when moving from random approach to the one of adding a few edges selectively. 

Figure \ref{fig:result3_build} illustrates that, in order to have a highly repairable network, it is not necessary to build the network from scratch; even a very small number of edges added using the proposed criteria, at the end of a randomly built network can elevate repairability to a level that is comparable to a network with the optimal repairability (evenly distributed edges of ring structure). Therefore, during the design process of a spare sharing network,  most edges can be added based on the requirement of the systems, and adding few edges in the last phase as suggested by the proposed methodology seems to be practical to enhance the system repairability significantly.

\begin{figure}[tb]
\centering
	\includegraphics[scale=.5]{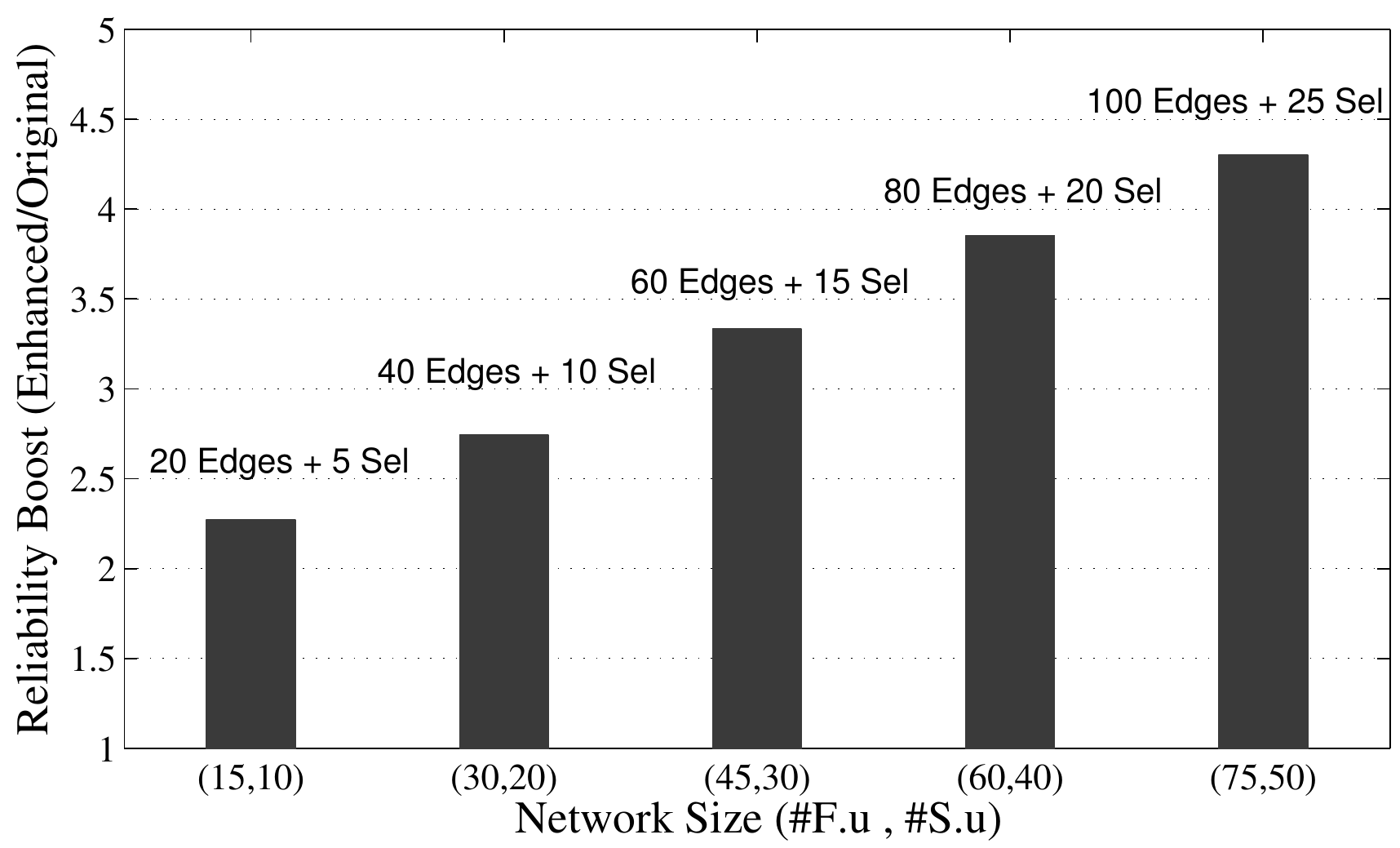}
		\caption{Scalability of the proposed methodology (the number of initial and extra edges for the first network is 20 and 5, respectively).}
	\label{fig:result2_scale}
\end{figure}

Figure \ref{fig:result2_scale} verifies the scalability of the proposed methodology, i.e., the larger the network size, the more gain can be achieved by the proposed methodology. It shows that the repairability boost of the enhanced network over the original network increases as system size scales. As can be seen in Figure \ref{fig:result2_scale}, by scaling the network size (the number of functional units, spare units, initial edges, and extra edges) the proposed approach can elevate the average repairability (over the faults) more significantly for the larger systems.

%--------------------------------------------------------------------------------------------------------------
%--------------------------------------------------------------------------------------------------------------
%--------------------------------------------------------------------------------------------------------------
%--------------------------------------------------------------------------------------------------------------
%--------------------------------------------------------------------------------------------------------------
%--------------------------------------------------------------------------------------------------------------
%--------------------------------------------------------------------------------------------------------------

\section{Conclusion}

We have studied the problem of boosting the repairability of future scalable systems with shared spares. Interconnection constraint is taken into consideration for scalable systems, such that the spares can only be shared in a limited way. First, we proposed several replacement algorithms, and showed how PE+PP achieves high repairability for both one system under worst-case scenario, and in overall repairability curve. Next, a low cost methodology is proposed to boost the repairability of any given network by adding a very small number of extra connections to expand spare sharing. We developed a set of criteria to pinpoint the most vulnerable functional units and the most exploitable spare units in the network, so that connecting them together will significantly increase the system repairability. Simulation results confirm that the proposed methodology is highly effective and cost-efficient in repairability boost for scalable systems. 

%\balance

\bibliographystyle{./IEEEtran}
\bibliography{./ss}

%\begin{IEEEbiographynophoto}{Soroush Khaleghi}
%received his B.Sc. from the Electrical Engineering department at Sharif University of Technology in Iran at %2007. He is currently a Ph.D. student at the department of Electrical and Computer Engineering at %University of Illinois at Chicago (UIC). His research is focused on the areas of hardware security and design for trust, as well as fault tolerance for future nanoelectronic systems. 
%\vfill

%\noindent
%\textbf{{Wenjing Rao}}
%received her B.Sc. in Computer Science from the CSE department at Beijing University in China. She received her Ph.D. in Computer Science from CSE department at University of California, San Diego (UCSD). She is currently an Associate Professor in the department of Electrical and Computer Engineering at University of Illinois at Chicago (UIC). Her research interests lie in the areas of nanoelectronic systems, reliability, fault and defect tolerance, digital test and design for testability, and VLSI CAD. 
%\vfill



\end{document}